\documentclass[a4paper,12pt]{article}

\usepackage{amsmath}\usepackage{amssymb}
\usepackage{latexsym}\usepackage{cite}

\usepackage[dvips]{graphicx}
\usepackage{pstricks}
\usepackage{bm}

\usepackage{graphicx}

\usepackage{amsmath, amssymb, graphics}

\newcommand{\mathsym}[1]{{}}

\newcommand{\be}{\begin{equation}}
\newcommand{\ee}{\end{equation}}

\newcommand{\ka}{\kappa}

\def\beq{\begin{equation}}
\def\eeq{\end{equation}}
\def\beqr{\begin{eqnarray}}
\def\eeqr{\end{eqnarray}}

\def\al{\alpha}
\def\bt{\beta}
\def\Ga{\Gamma}
\def\ga{\gamma}
\def\de{\delta}
\def\De{\Delta}

\def\ka{\kappa}
\def\si{\sigma}
\def\Si{\Sigma}
\def\te{\theta}

\def\lam{\lambda}

\def\om{\omega}
\def\ep{\epsilon}

\def\sq{\sqrt}

\def\l{\left (}
\def\r{\right )}

\def\fr{\frac}
\def\la{\label}
\def\hs{\hspace}
\def\vs{\vspace}

\def\ran{\rangle}
\def\lan{\langle}
\def\ov{\overline}

\def\tm{\times}

\begin{document}

\begin{flushright}
CETUP*12-019\\
March 6, 2013 \\
\end{flushright}

\vs{1.5cm}

\begin{center}
{\Large\bf

Three Family $SU(5)$ GUT and\\
Inverted Neutrino Mass Hierarchy}

\end{center}

\vspace{0.5cm}
\begin{center}
{\large
{}~Zurab Tavartkiladze\footnote{E-mail: zurab.tavartkiladze@gmail.com}
}
\vspace{0.5cm}

{\em Center for Elementary Particle Physics, ITP, Ilia State University, 0162 Tbilisi, Georgia}

\end{center}
\vspace{0.6cm}

\begin{abstract}


Supersymmetric $SU(5)$ GUT augmented with anomaly free $U(1)_F$ flavor symmetry is presented.
Very economical field content and $U(1)_F$ charge assignment are obtained by specific construction.
 In particular, three families of $10+\bar 5$ chiral matter, along the $SU(5)$ singlet
 states (some of which serve as right handed neutrinos) are obtained.
 Appealing texture zero Yukawa matrices provide natural understanding of hierarchies between charged fermion
 masses and mixings. The model predicts inverted hierarchical neutrino mass
 scenario with interesting implications.

\end{abstract}





\section{Introduction}

Although being very successful, the Standard Model  is unable to resolve some puzzles.
Among them is a problem of fermion flavor. The origin of hierarchies between charged fermion masses and CKM mixing
angles is unexplained. Is there any underlying theory which might generate these hierarchies in a natural way?
Moreover, in order to explain the neutrino data \cite{nu-data, An:2012eh, Schwetz:2011zk} some extension of the Standard Model,
generating neutrino masses and mixings, is necessary.
The number of fermion generations is a mystery. Do we have only three chiral families of quarks and leptons?
Is any selection rule dictating the  number of fermion generations?

Motivated by these questions,  in this paper we address these issues within the framework of supersymmetric (SUSY) $SU(5)$ Grand Unified Theory (GUT). Latter's motivation is to have unified description of electro-weak and strong interactions
\cite{Pati:1974yy}, while SUSY provides natural understanding of gauge hierarchy problem as well as the successful gauge coupling
unification \cite{Dimopoulos:1981yj}.
For understanding hierarchies between fermion masses and mixings, we apply Abelian flavor symmetry \cite{Froggatt:1978nt}
$U(1)_F$ which by requirement is non-anomalous. The $U(1)_F$, in combination with SUSY $SU(5)$ GUT, due to anomaly constraint
allows only three chiral families of matter $(10+\bar 5)$-plets and few $SU(5)$ singlet states. We use some of these singlet
states as right handed neutrinos, in order to build realist neutrino sector. Anomaly constraints fix $U(1)_F$ charge assignment
in such a way that texture zero quark and lepton Yukawa matrices are generated.
Together with  natural understanding of hierarchies, model predicts inverted neutrino mass
hierarchical scenario blending well with recent Daya Bay observation \cite{An:2012eh}.

\section{Three Family SUSY $SU(5)\tm U(1)_F$}

Consider SUSY $SU(5)$ GUT augmented with anomaly free $U(1)_F$ flavor symmetry.
Setup with anomaly free $U(1)_F$, will alow to gauge $U(1)_F$ and remain within conventional 4-dimensional field theoretical
framework (without need of discussing $U(1)$'s of a stringy origin \cite{Dine:1987xk}).
In a recent work \cite{Tavartkiladze:2011ex} the way of finding anomaly free $U(1)_F$ flavor symmetry within SUSY $SU(5)$ GUT was suggested.
The finding was realized by embedding of $SU(5)\tm U(1)_F$ in a single non-Abelian group with anomaly free field
content.\footnote{The way of this finding differs from those used earlier \cite{Dudas:1995yu}.}
In this way, the $U(1)_F$ charge assignment can be fixed. Amongst several assignments, found in \cite{Tavartkiladze:2011ex}, there is one which also dictates the number of generations to be three. This, As will be shown below, leads to very economical
and attractive scenario for fermion masses and mixings.
Before showing this, we briefly discuss the way of finding of such $U(1)_F$.

The states non-trivial under $SU(5)$ group, we introduce, will be just those of minimal SUSY $SU(5)$. These are scalar superfields
$\Si(24), H(5), \bar H(\bar 5)$ and three families of matter $(10+\bar 5)$ supermultiplets. We assume that $\Si$ is not charged under $U(1)_F$ and thus does not contribute to the anomalies. Therefore, upon finding anomaly free $U(1)_F$ charge assignment we will deal with three
$10$-plets, one $5$-plet (which is $H$) and four $\bar 5$-plets ($=$three matter $\bar 5$-plets plus $\bar H$). As already mentioned, we search
$U(1)_F$ charge assignment by embedding of $SU(5)\tm U(1)_F$ in non-Abelian group with anomaly free matter \cite{Tavartkiladze:2011ex}.
Let us consider $SU(7)$ group with chiral supermultiplets $35+2\tm \bar 7$. This simple set is anomaly free \cite{anomalies}.
Here $35$ is three index antisymmetric representation and $\bar 7$ is an anti-fundamental of $SU(7)$.
Decomposition of these states via the chain
$SU(7)\to SU(6)\tm U(1)_7\to SU(5)\tm U(1)_6\tm U(1)_7$ looks
$$
35=20_3+15_{-4}=(10_{-3}+\ov{10}_3)_3+(10_2+5_{-4})_{-4}~,
$$
\beq
\bar 7=\bar 6_{-1}+1_6=(\bar 5_{-1}+1_5)_{-1}+(1_0)_6~,
\la{SU7}
\eeq
where subscripts inside and outside of parenthesis indicate $U(1)_6$ and $U(1)_7$ charges respectively. Note that $U(1)_6$ and $U(1)_7$ are coming
from $SU(6)$ and $SU(7)$ respectively, with corresponding generators  $Y_{U(1)_6}=\fr{1}{\sqrt{60}}{\rm Diag}\l 1, 1, 1, 1, 1, -5\r $ and
$Y_{U(1)_7}=\fr{1}{\sqrt{84}}{\rm Diag}\l 1, 1, 1, 1, 1, 1, -6\r $. The normalization factors $\fr{1}{\sqrt{60}}$   and $\fr{1}{\sqrt{84}}$
are omitted in Eq. (\ref{SU7}).
Now, in (\ref{SU7}), without change of charge assignments, we replace the pair of $(\ov{10}+5)$-plets by the pair $(10+\bar 5)$.
With this replacement all anomalies (at the level of $SU(5)$ and $U(1)$'s) will remain intact, i.e. will still vanish. Thus, we will have the following anomaly free
content
\beq
(10_{-3}+10_3)_3+(10_2+\bar 5_{-4})_{-4}+2\tm \left [(\bar 5_{-1}+1_5)_{-1}+(1_0)_6 \right ]~,
\la{SU7-content}
\eeq
which involves three families (!) of matter $(10+\bar 5)$ supermultiplets plus four $SU(5)$ singlets. Some of these singlets
will be applied as right handed neutrinos (RHN). Worth noting that, in difference from $SO(10)$ GUT, the $SU(5)$ does not
involve (require) the RHN states. In the $SU(5)$ scenario, we have just built up, the RHNs are required for anomaly cancellation.

By Abelian symmetries  $U(1)_6$ and $U(1)_7$, with charges given in
 Eq. (\ref{SU7-content}), we can build superposition
$\bar aQ_{U(1)_6}+\bar bQ_{U(1)_7}$.
This superposition is automatically anomaly free for arbitrary $\bar{a}$ and $\bar{b}$, because the orthogonal generators $Y_{U(1)_6}$ and $Y_{U(1)_7}$
originate from single $SU(7)$. Thus, using (\ref{SU7-content}) we can write the anomaly free set
\beq
10_{-3\bar{a}+3\bar{b}}+10_{3\bar{a}+3\bar{b}}+10_{2\bar{a}-4\bar{b}}+\bar 5_{-4\bar{a}-4\bar{b}}+
2\tm \left (\bar 5_{-\bar{a}-\bar{b}}+1_{5\bar{a}-\bar{b}}+1'_{6\bar{b}}  \right )~,
\la{supSU6-SU7}
\eeq
where subscripts denote  charges.
As it turns out, for building realistic phenomenology it is useful to  add to this superposition another anomaly free $U(1)$,
which can be found by similar procedure. For instance, consider
$27$-plet of $E_6$ group with a chain $E_6\to SO(10)\tm U(1)_{E_6}\to SU(5)\tm U(1)_{E_6}$ of decomposition
\cite{Slansky:1981yr}:
\beq
27=16_1+10_{-2}+{1'}_{4}=(10+\bar 5+1)_1+(5+\bar 5')_{-2}+{1'}_{4}~.
\la{flipSO10}
\eeq
Here subscripts denote $U(1)_{E_6}$ charges. In this content, we can replace $5_{-2}$ with $\bar 5_{-2}$
and at the same time add $10_p+10_{-p}$. Moreover, we can add two  $SU(5)$ singlets with $U(1)_{E_6}$ charges $k$ and $-k$ respectively.
With this replacement and additions, the anomalies $SU(5)^3$, $SU(5)^2\cdot U(1)_{E_6}$, etc, will be unchanged. This content allows to build superposition
of three Abelian groups $U(1)_6$, $U(1)_7$ and $U(1)_{E_6}$:
$\bar Q_{sup}=\bar aQ_{U(1)_6}+\bar bQ_{U(1)_7}+\bar cQ_{U(1)_{E_6}}$. Thus, the field content and
$\bar Q_{sup}$ charge assignment will look:
$$
10_{-3\bar{a}+3\bar{b}+p\bar{c}}+10_{3\bar{a}+3\bar{b}-p\bar{c}}+10_{2\bar{a}-4\bar{b}+\bar{c}}+\bar 5_{-4\bar{a}-4\bar{b}-2\bar{c}}+
 \bar 5_{-\bar{a}-\bar{b}+\bar{c}}+\bar 5_{-\bar{a}-\bar{b}-2\bar{c}}
 $$
 \beq
 +1_{5\bar{a}-\bar{b}+\bar{c}}+1_{5\bar{a}-\bar{b}+4\bar{c}}+1'_{6\bar{b}+k\bar{c}}+1'_{6\bar{b}-k\bar{c}}~,
 ~~~~~{\rm with}~~~~30\bar{a}(3+2p)=\bar{c}(2k^2+10p^2-27)~.
\la{D-prime}
\eeq
Relations between $\bar{a}, \bar{c}, k$ and $p$ (imposed for  $\bar{c}\neq 0$) given in Eq. (\ref{D-prime}) insures that all anomalies vanish. Clearly, with rational selection of $\bar{a}, k$ and $p$ the  value of $\bar{c}$
also will be rational. The set given in Eq. (\ref{D-prime}) is one simple selection among several options and opens up many possibilities for model building with realistic phenomenology. We will identify
these  charges with the charges of  $U(1)_F$ flavor symmetry.

\section{Model: Quark and Charged Lepton Yukawa Textures}

To the field content of Eq. (\ref{D-prime}) we add the pair $5_q+\bar 5_{-q}$. This is needed to have, besides
the matter fields, the Higgs supermultiplets $H+\bar H$. Thus, in total we have three $10$-plets,
one $5$-plet and four $\bar 5$-plets plus $SU(5)$ singlets. The $U(1)_F$ charge assignment ($q, -q$ for $5_q, \bar 5_{-q}$
and for remaining $\bar 5$-plets given in Eq. (\ref{D-prime})) is not unique. We can exchange $5$-plet's $U(1)_F$ charge
with one of the $\bar 5$-plets' charge. With this, all anomalies will still vanish. In addition, out of the four $\bar 5$-plets,
any of them can be identified with the Higgs superfield $\bar H$. As it turns out, for the charges of the
pair  $(H, \bar H)$, we will have $13$ possible options for $(Q_H, Q_{\bar H})$:
\beqr
\hs{-0.5cm}(Q_H, Q_{\bar H})^{(l)}&\hs{-0.5cm}=&\hs{-0.4cm} \left \{(q, -q), ~(q, -4\bar{a}-4\bar{b}-2\bar{c}),~ (q, -\bar{a}-\bar{b}+\bar{c}), ~
(q, -\bar{a}-\bar{b}-2\bar{c}), \right.  \nonumber \\
&& \hs{-0.2cm}(-4\bar{a}-4\bar{b}-2\bar{c}, -\bar{a}-\bar{b}+\bar{c}), (-4\bar{a}-4\bar{b}-2\bar{c}, -\bar{a}-\bar{b}-2\bar{c}), (-4\bar{a}-4\bar{b}-2\bar{c}, q),
 \nonumber \\
&& \hs{-0.2cm}(-\bar{a}-\bar{b}+\bar{c}, -4\bar{a}-4\bar{b}-2\bar{c}),(-\bar{a}-\bar{b}+\bar{c}, -\bar{a}-\bar{b}-2\bar{c}),(-\bar{a}-\bar{b}+\bar{c},q), \nonumber \\
&&\left.  \hs{-0.2cm} (-\bar{a}-\bar{b}-2\bar{c},-4\bar{a}-4\bar{b}-2\bar{c}),(-\bar{a}-\bar{b}-2\bar{c},-\bar{a}-\bar{b}+\bar{c}),(-\bar{a}-\bar{b}-2\bar{c},q)
\right\},
\la{D-QHHbar}
 \eeqr
with $l=1,\cdots, 13$. Note that we have left out possibilities obtained from those given in (\ref{D-QHHbar}) by the substitution $q\to -q$. These $13$
options open up various possibilities for the model building \cite{inprep}.
 Below we present one of them, which we found to have nice and attractive properties with interesting implications for fermion masses and mixings.

\subsubsection*{$U(1)_F$ symmetry breaking}

In order to break $U(1)_F$ gauge symmetry, we introduce $SU(5)$ singlet pair of flavon superfields $X+\bar X$
with $U(1)_F$ charges
\beq
Q(X)=-1 ~,~~~~~~~~Q(\bar X)=1~.
\la{XbarX-ch} 
\eeq
Without loss of generality, we have normalized flavons' charges modulo to one. The scalar components of
 $X$ and $\bar X$ acquire
 VEVs\footnote{Details of symmetry breaking, with general values of $ \lan X\ran$ and $\lan \bar X\ran $, is given in
Ref. \cite{Tavartkiladze:2011ex}.}
\beq
\fr{\lan |X|\ran }{M_{\rm Pl}}= \ep ~,~~~~\fr{\lan |\bar X|\ran }{M_{\rm Pl}}= \bar{\ep }~,
\la{eps-epsbar}
\eeq
where $M_{\rm Pl}\simeq 2.4\cdot 10^{18}$~GeV is reduced Planck scale, which will be treated as natural cut off for all higher dimensional non-renormalizable operators.
In our approach, top quark (and possibly bottom quark and tau lepton, in case of large $\tan \bt $) will get mass
 at renormalizable level. Yukawa couplings of light families emerge after $U(1)_F$ flavor symmetry breaking.
Thus, the hierarchies between Yukawa couplings and CKM mixing angles will be expressed by powers of small parameters
$\ep, \bar{\ep }\ll 1$.

\subsubsection*{Yukawa textures}

In the charge assignment we need to fix the values of $\bar a, \bar b$ and $\bar c$. If their ratios remain arbitrary there will be more than one extra $U(1)$ symmetry, and that we have to avoid. Together with fixing  $\bar a, \bar b, \bar c$, the
values of $p, q$ and $k$ (in (\ref{D-prime}) and (\ref{D-QHHbar})) should be selected in such a way as to have
phenomenologically viable  quark and lepton Yukawa textures.
It turns out, that one selection leading to attractive Yukawa sector, is the following:
\beq
\{ \bar a, \bar b, \bar c\}=\left \{-\fr{1}{2}, \fr{1}{6}, \fr{5}{3} \right \}~, ~~~~~~~~p=q=k=0~.
\la{selcts-abc-pqk}
\eeq
In this case, in Eq. (\ref{D-QHHbar}) we pick up $l=3$, which fixes charges of $H, \bar H$ as
$(Q_H, Q_{\bar H})^{l=3}=(0, 2)$. The charges of matter $(10+\bar 5)$-plets (and also $SU(5)$ singlets) are also fixed
and we will make the following identification:
$Q_{10_i}=\{ 2, -1, 0\}~,~Q_{\bar 5_i}=\{0, -3, -2 \}$, where $i=1,2,3$ labels the flavor.
The model's field content and corresponding $U(1)_F$ charges are given in Table \ref{t:tab1}.
%
%
%
\begin{table}
\caption{$U(1)_F$ charge assignment for the model's states.
 }

\label{t:tab1} $$\begin{array}{|c||c|c|c|c|c|c|c|c|c|c|c|c|c|c|c|}

\hline
\vs{-0.3cm}
 &  &  &  &  &  &  &  &&& & &&&&\\

\vs{-0.4cm}

& \hs{0.3mm}10_1\hs{0.3mm}& \hs{0.3mm}10_2\hs{0.3mm}&\hs{0.3mm}10_3\hs{0.3mm}& \hs{0.5mm}\bar 5_1 \hs{0.5mm}&
\hs{0.5mm}\bar 5_2\hs{0.5mm} &\hs{0.5mm}\bar 5_3\hs{0.5mm} &\hs{-0.5mm}H(5) \hs{-0.5mm}&\hs{-0.5mm} \bar H(\bar 5)\hs{-0.5mm}  & \hs{-0.5mm}\Si (24)\hs{-0.5mm}&X &\bar X&~1_1 ~ & ~1_2 & 1_3&1_4\\

&  &  &  &  &  &  &  &  & &&&&&&\\

\hline

\vs{-0.3cm}
 &  &  &  &  &  &  &  & & &&&&&&\\

\vs{-0.3cm}
\hs{-0.5mm}Q_{U(1)_F}\hs{-0.5mm}& 2 & -1 & 0 & 0 & -3 & -2 &0 &2 &0 &-1 &1&1&1 &-1 &4\\

&  &  &  &  &  &  &  & & &&&&&&\\

\hline

\end{array}$$

\end{table}
%
%
%
With this assignment, $10\cdot 10H$ and $10\cdot \bar 5\bar H$-type Yukawa couplings are given by
\beq
\begin{array}{ccc}
 & {\begin{array}{ccc}
\hs{-0.9cm}10_1 &10_2  &~10_3 \hs{0.2cm}
\end{array}}\\ \vspace{1mm}
\begin{array}{c}
 10_1\\ 10_2 ~  \\ 10_3 ~
 \end{array}\!\!\!\!\!\hs{-0.1cm}&{\!\left(\begin{array}{ccc}

 \ep^4& ~~\ep  & ~~\ep^2
\\
 \ep &~~ \bar{\ep}^{\hs{0.1cm}2} & ~~ \bar{\ep}
 \\
\ep^2& ~~ \bar{\ep}  &~~ 1
\end{array}\right)H}~,~~~
\end{array}  \!\!  ~~~
\begin{array}{ccc}
 & {\begin{array}{ccc}
\hs{-0.6cm}\bar 5_1 &~~ \bar 5_2  &~~\bar 5_3 \hs{0.2cm}
\end{array}}\\ \vspace{1mm}
\begin{array}{c}
 10_1\\ 10_2 ~  \\ 10_3 ~
 \end{array}\!\!\!\!\!\hs{-0.1cm}&{\!\left(\begin{array}{ccc}

 \ep^4& ~~\ep  & ~~\ep^2
\\
 \ep &~~ \bar{\ep}^2 & ~~\bar{\ep}
 \\
\ep^2& ~~ \bar{\ep}  &~~ 1
\end{array}\right)\bar H}~,
\end{array}  \!\!
\label{UDE-d-u2}
\eeq
where in front of each entry, dimensionless couplings($\sim 1/5-5$) are assumed. As we will see shortly,
good fit is achieved for $\bar{\ep}\sim 1/10~,~\ep \sim (0.05-0.2)\bar{\ep}^2$.
On the other hand, with these values, the matrix elements
$(1,1), (1,3), (3,1)$ are so suppressed, that they are irrelevant and we can set them equal to zero.
Thus, for all practical purposes, we can investigate the Yukawa matrices:
\beq
\begin{array}{ccc}
 & {\begin{array}{ccc}
\hs{-0.6cm} &~~  &~~ \hs{0.2cm}
\end{array}}\\ \vspace{1mm}
\begin{array}{c}
 \\  ~  \\  ~
 \end{array}\!\!\!\!\!\hs{-0.1cm}&{\!Y_{U,D,E}\propto \left(\begin{array}{ccc}

 0& ~~\ep  & ~~0
\\
 \ep &~~ \bar{\ep}^2 & ~~\bar{\ep}
 \\
0& ~~ \bar{\ep}  &~~ 1
\end{array}\right)}~,
\end{array}  \!\!  ~~~
\label{texture-0}
\eeq
with zero textures.

\subsubsection*{\bf Quark masses and mixings}

Using the basis $q^TY_Uu^ch_u$ and $q^TY_Dd^ch_d$, without loss of generality we can parameterize
up and down Yukawa matrices at GUT scale to have forms:
\beq
 Y_U\simeq \left(\begin{array}{ccc}

 0& ~~ c\lam \bar{\ep }^{\hs{0.1cm}2}& ~~ 0
\\
 c\lam \bar{\ep }^{\hs{0.1cm}2} &~~ a_u\bar{\ep }^{\hs{0.1cm}2}e^{i\xi_u} & ~~ \bar{\ep }
 \\
0 & ~~ \bar{\ep }  &~~ 1
\end{array}\right)\lam_t^0~,
\label{UpY-DU2}
\eeq
\beq
 Y_D\simeq
 \left(\begin{array}{ccc}

 e^{i\varphi'}& ~~ 0& ~~ 0
\\
 0 &~~ e^{i\varphi} & ~~ 0
 \\
0 & ~~ 0 &~~ 1
\end{array}\right)
\left(\begin{array}{ccc}

 0& ~~ \lam \bar{\ep }^{\hs{0.1cm}2}& ~~ 0
\\
 k\lam \bar{\ep }^{\hs{0.1cm}2} &~~ a_d\bar{\ep }^{\hs{0.1cm}2}e^{i\xi_d} & ~~ b\bar{\ep }
 \\
0 & ~~ b'\bar{\ep }  &~~ 1
\end{array}\right)\lam_b^0~.
\label{DownY-DU2}
\eeq
We have made field phase redefinitions in such a way that, in this basis, CKM matrix remains unity and in $Y_U$ only one phase $\xi_u$ appears. The phases $\varphi$ and $\varphi'$ will not contribute to the quark masses, but will be important for the CKM matrix elements.
From  Eqs. (\ref{UpY-DU2}) and (\ref{DownY-DU2}), in a fairly good approximation  we obtain the following relations
valid at GUT scale:
\beq
\fr{\lam_c}{\lam_t}\simeq \bar{\ep }^{\hs{0.1cm}2}\left |a_u\bar{\ep }^{\hs{0.1cm}2}e^{i\xi_u}-1\right |~,~~~
c\lam \simeq \sqrt{\fr{\lam_u}{\lam_c}}\left |a_u\bar{\ep }^{\hs{0.1cm}2}e^{i\xi_u}-1\right |~,
\la{up-asymp}
\eeq
\beq
\fr{\lam_s}{\lam_b}\simeq \bar{\ep }^{\hs{0.1cm}2}\fr{\left | a_d\bar{\ep }^{\hs{0.1cm}2}e^{i\xi_d}-bb'\right |}{1+(b'\bar{\ep })^2}~,~~~
\lam \bar{\ep }^{\hs{0.1cm}2}\sqrt{k}\simeq \sqrt{\fr{\lam_d\lam_s}{\lam_b^2}}\l 1+(b'\bar{\ep })^2\r^{3/4}.
\la{down-asymp}
\eeq
These relations help to find a good fit.
With  proper selection of input parameters $\bar{\ep}, \lam $, $a_{u, d}$, $b, b', c, c', k$, $\xi_{u, d}$, $\varphi, \varphi'$ we can get desirable values for fermion mass hierarchies and CKM mixing angles at GUT scale.
Then, using RG we can calculate these ratios at low scales:
$$
\left. \fr{\lam_{u,c}}{\lam_t}\right |_{m_t}=\eta_t^3\eta_b\left. \fr{\lam_{u,c}}{\lam_t}\right |_{M_G}~,~~~
\left. \fr{\lam_{d,s}}{\lam_b}\right |_{M_Z}=\eta_t\eta_b^3\left. \fr{\lam_{d,s}}{\lam_b}\right |_{M_G}~,~~~
\left. \fr{\lam_{e,\mu}}{\lam_{\tau}}\right |_{M_Z}=\eta_{\tau}^3\left. \fr{\lam_{e,\mu}}{\lam_{\tau}}\right |_{M_G}~,~~~
$$
$$
\left. V_{\al \bt}\right |_{M_Z}=\eta_t\eta_b \left. V_{\al \bt}\right |_{M_G}~,~~~{\rm if}~~~(\al \bt )=(ub, cb, td, ts)
$$
\beq
\left. V_{\al \bt}\right |_{M_Z}=\left. V_{\al \bt}\right |_{M_G}~,~~~{\rm if}~~~(\al \bt )=(ud, us, cd, cs, tb)~,
\la{RG-MG-mt}
\eeq
where RG factors
\beq
\eta_t=\exp \!\l \!\fr{1}{16\pi^2}\!\!\int_{m_t}^{M_G}\hs{-0.2cm}\lam_t^2d \ln \mu \!\r ,~~
\eta_b=\exp \!\l \!\fr{1}{16\pi^2}\!\!\int_{m_h}^{M_G}\hs{-0.2cm}\lam_b^2d \ln \mu \!\r ,~~
\eta_{\tau}=\exp \!\l \!\fr{1}{16\pi^2}\!\!\int_{m_h}^{M_G}\hs{-0.2cm}\lam_{\tau }^2d \ln \mu \!\r
\la{RG-int}
\eeq
are given in 1-loop approximation.

We will consider two cases with low/moderate and large values of the MSSM parameter $\tan \beta $.

\subsubsection*{Fit for $\tan \bt =5-15$}

We take experimental value $m_t(m_t)=163.68$~GeV, determining top Yukawa coupling at weak scale,
and with $\tan \bt =5-15$ we find $\eta_t=1.097, \eta_b\simeq \eta_{\tau }\simeq 1$~.
For this case, good fit is obtained for the following values of input parameters:
$$
\bar{\ep }=0.0847 ~,~~\lam =0.476 ~,~~ a_u=0.6 ~,~~ a_d=3.7 ~,
$$
$$
b=-0.798~,~~b'=-7.14 ~, ~~c=0.037 ~,~~k=0.864 ~,
$$
\beq
\xi_u=0 ~,~~\xi_d=-0.065 ~,~~\varphi =-2.696 ~,~~\varphi'=-0.97 ~.
\la{inp-tan5}
\eeq
These at GUT scale give
$$
{\rm at}~~\mu=M_G:~~~\fr{\lam_u}{\lam_t}=5.51\cdot 10^{-6} ~,~~\fr{\lam_c}{\lam_t}=0.002835 ~,~
~\fr{\lam_d}{\lam_b}=5.64\cdot 10^{-4} ~,~~\fr{\lam_s}{\lam_b}= 0.0111~,~~
$$
$$
|V_{us}|=0.2243~,~~|V_{cb}|=0.0383~,~~|V_{ub}|=0.00318~,~~~\ov{\rho}=0.118 ~,~~\ov{\eta}=0.34~,
$$
where $\ov{\rho}+i\ov{\eta}=-\fr{V_{ud}V_{ub}^*}{V_{cd}V_{cb}^*}$~.

Performing renormalization (using (\ref{RG-MG-mt}) and \cite{Fusaoka:1998vc}), at low scales we get (with input $m_t(m_t)=163.68$~GeV, $m_b(m_b)=4.24$~GeV):
 $$
 \l m_u,~ m_d,~ m_s,~ m_c\r(2~{\rm GeV})=\l 2.1,~ 4.64, ~91.69,~ 1082\r {\rm MeV}
 $$
 \beq
 {\rm at}~~\mu=M_Z:~~|V_{us}|=0.2243~,~~|V_{cb}|=0.042~,~~|V_{ub}|=0.00349~,~~~\ov{\rho}=0.118 ~,~~\ov{\eta}=0.34~.
 \la{low-sc}
 \eeq
These values of masses and CKM matrix elements are in good agreement with experiments \cite{Davies:2009ih}, \cite{Nakamura:2010zzi}.

\subsubsection*{Fit for $\tan \bt =55$}

In this case we have
$\eta_t=1.114, \eta_b=1.158,~ \eta_{\tau }=1.105$~.
Input parameters are selected as:
$$
\bar{\ep }=0.0723 ~,~~\lam =0.53 ~,~~ a_u=0.545 ~,~~ a_d=6.81 ~,
$$
$$
b=-0.777~,~~b'=-11.58 ~, ~~c=0.0375 ~,~~k=0.783 ~,
$$
\beq
\xi_u=-0.055 ~,~~\xi_d=-0.0593 ~,~~\varphi =-2.73 ~,~~\varphi'=-0.98 ~,
\la{inp-tan55}
\eeq
giving at GUT scale
$$
{\rm at}~~\mu=M_G:~~~\fr{\lam_u}{\lam_t}=4.5\cdot 10^{-6} ~,~~\fr{\lam_c}{\lam_t}=0.002367 ~,~
~\fr{\lam_d}{\lam_b}=3.73\cdot 10^{-4} ~,~~\fr{\lam_s}{\lam_b}= 0.00723~,~~
$$
$$
|V_{us}|=0.2259~,~~|V_{cb}|=0.0317~,~~|V_{ub}|=0.00272~,~~~\ov{\rho}=0.135 ~,~~\ov{\eta}=0.345~.
$$
The renormalization procedure (using (\ref{RG-MG-mt}) and \cite{Fusaoka:1998vc}) gives at low scales (with input $m_t(m_t)=163.68$~GeV, $m_b(m_b)=4.24$~GeV):
 $$
 \l m_u,~ m_d,~ m_s,~ m_c\r(2~{\rm GeV})=\l 2.08,~ 4.85, ~93.86,~ 1096\r {\rm MeV}
 $$
 \beq
 {\rm at}~~\mu=M_Z:~~|V_{us}|=0.2259~,~~|V_{cb}|=0.0409~,~~|V_{ub}|=0.00351~,~~~\ov{\rho}=0.135 ~,~~\ov{\eta}=0.345~.
 \la{low-sc}
 \eeq
These agree well with experiments.

\subsubsection*{Charged lepton sector}

Now let us discuss the charged lepton sector. Relevant Yukawa couplings
originate from $10\cdot \bar 5\cdot \bar H$-type interactions of Eq. (\ref{UDE-d-u2}) (while in practice $Y_E$
has the structure of Eq. (\ref{texture-0})). Without breaking the
$SU(5)$ symmetry in these interactions, one would get the asymptotic relation $M_D=M_E^T$, which is
unacceptable and is a well known problem for minimal SUSY $SU(5)$ GUT.
However, by some specific extension, care can be exercised to solve this
problem \cite{Babu:2012pb}. Without specifying origin
of $SU(5)$ breaking in this sector,
we assume that it happens (i.e. $SU(5)$ symmetry breaking) in the sector of light families.
Thus, in analogy of $Y_D$ (see Eq. (\ref{DownY-DU2})), in a basis $l^TY_Ee^ch_d$, we parameterize $Y_E$ to have the
following form
\beq
 Y_E\simeq \left(\begin{array}{ccc}

 0& ~~ k_e\lam_e \bar{\ep }^{\hs{0.1cm}2}& ~~ 0
\\
 \lam_e \bar{\ep }^{\hs{0.1cm}2} &~~ k_{22}a_d\bar{\ep }^{\hs{0.1cm}2}e^{i\xi_e} & ~~ b'\bar{\ep }
 \\
0 & ~~ b\bar{\ep }  &~~ 1
\end{array}\right)\lam_{\tau }^0~,~~~~~{\rm with}~~~~~\lam_{\tau }^0=\lam_b^0~.
\label{YE-DU2}
\eeq
In $Y_E$ only one complex phase $\xi_e$ appears. Remaining phases are rotated away by proper phase redefinitions of
the $l$ and $e^c$ states.
With $\{k_{22}, k_e, \lam_e, \xi_e\}\neq \{1, k, \lam, \xi_d \}$ we can avoid the relation $\fr{m_d}{m_s}=\fr{m_e}{m_{\mu }}$, while keeping $m_b^0=m_{\tau }^0$ (at the GUT scale).
Good fit can be obtained with
$$
{\rm for}~~\tan \bt =5-15~,~~~~\lam_e=2.51~,~~~~k_e=0.082~,~~~~k_{22}=4.517~,~~\xi_e=-0.065,~~
$$
\beq
{\rm for}~~\tan \bt =55~,~~~~~~~~~\lam_e=3.11~,~~~~k_e=0.07656~,~~~~k_{22}=3.3677~,~~\xi_e=-0.06~,~~
\la{E-pars}
\eeq
and remaining parameters given in Eqs. (\ref{inp-tan5}) and (\ref{inp-tan55}) respectively. With these we obtain
$$
{\rm at}~~\mu=M_G~,~~~~{\rm for}~~\tan \bt =5-15~,~~~~\fr{\lam_e}{\lam_{\tau }}= 2.787\cdot 10^{-4}~,~~~~
\fr{\lam_{\mu }}{\lam_{\tau }}=0.05883 ~,
$$
\beq
{\rm at}~~\mu=M_G~,~~~~{\rm for}~~\tan \bt =55~,~~~~~~~~\fr{\lam_e}{\lam_{\tau }}=2.065\cdot 10^{-4} ~,~~~~
\fr{\lam_{\mu }}{\lam_{\tau }}=0.0436~.
\la{E-GUT}
\eeq
These lead to
\beq
m_e(m_e)= 0.511~{\rm MeV},~~~~m_{\mu }(m_{\mu })=105.66~{\rm MeV} ~,~~~~m_{\tau }(m_{\tau })=1.777~{\rm GeV} ~,
\la{lept-mass-low}
\eeq
in agreement with experiments.
The mixing angles originating from the charged lepton sector, for $\tan \bt =5-15$, are
$\{\te_{23}^e, \te_{12}^e, \te_{13}^e\} \simeq \{31.5^o, 1.02^o, 0.61^o\}$. While for
$\tan \bt =55$ we got $\{\te_{23}^e, \te_{12}^e, \te_{13}^e\} \simeq \{40.2^o, 0.93^o, 0.78^o\}$.
Note that while $\te_{23}^e$ is large (but not sufficiently), the  $\te_{12}^e$ and $\te_{13}^e$ are too small. This means that neutrino sector should be responsible for generating proper values of the lepton mixing angles.

\section{Neutrino Sector}
\la{nu-sect}

To build the realistic neutrino sector, we apply the singlet states ${\bf 1}_{1,2,3}$ (with $U(1)_F$
charges given in Table \ref{t:tab1}) as right handed neutrinos. Their Dirac type couplings (to $\bar 5_i$ states)
and the mass matrix respectively are given by:
\beq
\begin{array}{ccc}
 & {\begin{array}{ccc}
\hs{-1cm}{\bf 1}_1 &~~~{\bf 1}_2  &~~~{\bf 1}_3
\end{array}}\\ \vspace{1mm}
m_D\propto\begin{array}{c}
 \bar 5_1\\ \bar 5_2  \\ \bar 5_3
 \end{array}\!\!\!\!\!\hs{-0.1cm}&{\! \left(\begin{array}{ccc}

\ep & ~~\ep  & ~~\bar{\ep }
\\
~\bar{\ep}^{\hs{0.1cm}2} &~~ \bar{\ep}^{\hs{0.1cm}2} & ~~\bar{\ep}^{\hs{0.1cm}4}
 \\
\bar{\ep }& ~~ \bar{\ep}  &~~ \bar{\ep}^{\hs{0.1cm}3}
\end{array}\right)H}~,~~~
\end{array}  \!\!  ~~~
\begin{array}{ccc}
 & {\begin{array}{ccc}
\hs{-0.9cm}{\bf 1}_{1} &~~~{\bf 1}_{2}  &~~~{\bf 1}_{3}
\end{array}}\\ \vspace{1mm}
M_R\propto\begin{array}{c}
 {\bf 1}_{1}\\ {\bf 1}_{2}  \\ {\bf 1}_{3}
 \end{array}\!\!\!\!\!\hs{-0.1cm}&{\!\left(\begin{array}{ccc}

 \ep^2& ~~~\ep^2  & ~~0
\\
 \ep^2 &~~~ \ep^2 & ~~1
 \\
0& ~~1  &~~~  \bar{\ep}^2
\end{array}\right)M_{*}}~,
\end{array}
\label{Nu-coupl}
\eeq
where $M_{*}$ is some mass scale and in the entries of these matrices the dimensionless couplings are omitted.
Integration of heavy ${\bf 1}_i$ states leads to $3\tm 3$ mass matrix for the light neutrinos:
\beq
\begin{array}{ccc}
 & {\begin{array}{ccc}
\hs{-0.6cm} &~~  &~~ \hs{0.2cm}
\end{array}}\\ \vspace{1mm}
\begin{array}{c}
 \\  ~  \\  ~
 \end{array}\!\!\!\!\!\hs{-0.1cm}&{\! M_{\nu }=m_DM_R^{-1}m_D^T\stackrel{\propto }{_\sim } \left(\begin{array}{ccc}

 \ep^2& ~~\ep  \bar{\ep}^2 & ~~\ep \bar{\ep}
\\
 \ep  \bar{\ep}^2 &~~ \al^2\bar{\ep}^4 & ~~\al \bt \bar{\ep}^3
 \\
\ep \bar{\ep}& ~~\al \bt  \bar{\ep}^3 &~~ \bt^2\bar{\ep}^2
\end{array}\right)\bar m}~,
\end{array}  \!\!  ~~~
\label{nu-matrix1}
\eeq
with $\bar m\sim \fr{\lan h_u^{(0)}\ran^2}{M_{*}\ep^2}$ and $\al ,\bt $ are some dimensionless couplings. Note that,
$M_{\nu }$'s $2-3$ block's determinant is zero. It is convenient to work in
a basis where charged lepton mass matrix is diagonal, i.e. rotate whole lepton  doublets by unitary matrix which diagonalizes
the matrix $Y_EY_E^\dag $. In this basis, the weak leptonic current is diagonal and the neutrino mass matrix can be
denoted by $\bar M_{\nu}$.
 The convenience of this basis is that the diagonalizing matrix $U$:
\beq
U^T\bar M_{\nu }U=M_{\nu}^{\rm Diag}
\la{barMnu}
\eeq
will coincide with the lepton mixing matrix. The latter, in a standard parametrization, has the form:
\begin{equation}
U=
P_1\left(\begin{array}{ccc}c_{13}c_{12}&c_{13}s_{12}&s_{13}e^{-i\delta}\\
-c_{23}s_{12}-s_{23}s_{13}c_{12}e^{i\delta}&c_{23}c_{12}-s_{23}s_{13}s_{12}e^{i\delta}&s_{23}c_{13}\\
s_{23}s_{12}-c_{23}s_{13}c_{12}e^{i\delta}&-s_{23}c_{12}-c_{23}s_{13}s_{12}e^{i\delta}&c_{23}c_{13}
\end{array}\right)P_2
\la{Ulept}
\end{equation}
with $s_{ij}=\sin\theta_{ij}$ and $c_{ij}=\cos\theta_{ij}$. The phase matrices $P_{1,2}$ are given by:
\beq
P_1={\rm Diag}\l e^{-i\om_1}~,~e^{-i\om_2}~,~e^{-i\om_3}\r ~,~~~
P_2={\rm Diag}\l 1~,~e^{-i\rho_1/2}~,~e^{-i\rho_2/2}\r ~,
\la{Ps}
\eeq
where $\om_{1,2,3}, \rho_{1,2}$ are some phases.

\begin{figure}
\begin{center}
\leavevmode
\leavevmode
\vspace{-0.5cm}
\includegraphics{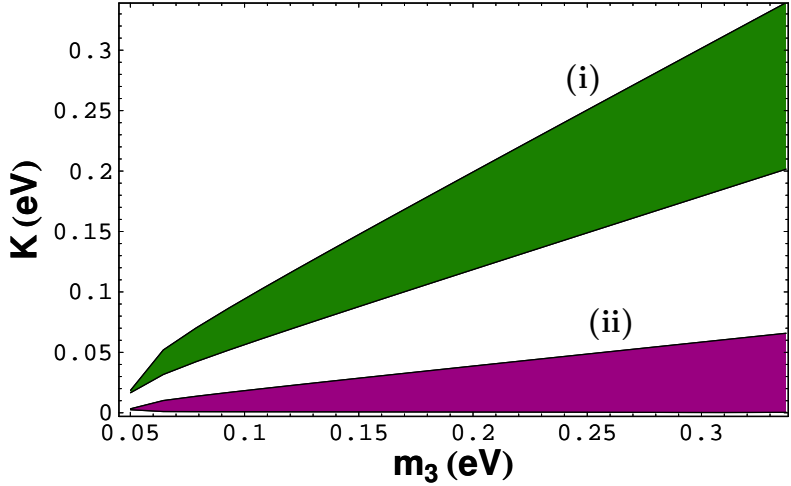}  
\end{center}
\vs{6.8cm}
\caption{Region {\bf (i)}: Needed values of $K$,
realizing normal hierarchical neutrino masses.
Region {\bf (ii)}: Values of $K$ within considered scenario with normal
ordering of neutrino masses of Eq. (\ref{norm-masses}).}
\rput(11.1,4.25){ \bf (ii)}
\rput(10.8,7.5){\bf (i)}
\label{fig1}
\end{figure}

To get some feeling about the results, obtained from the neutrino mass matrix, let us first
ignore $\te_{12}^e$ and $\te_{13}^e$ mixings. Since these angles are small, the picture qualitatively will remain unchanged.
(Effects of these mixing angles are discussed in detail in an Appendix).
With  $\te_{12}^e, \te_{13}^e\ll 1$, in a good approximation
$\bar M_{\nu}$  can be written as:
\beq
\begin{array}{ccc}
 & {\begin{array}{ccc}
\hs{-0.6cm} &~~  &~~ \hs{0.2cm}
\end{array}}\\ \vspace{1mm}
\begin{array}{c}
 \\  ~  \\  ~
 \end{array}\!\!\!\!\!\hs{-0.1cm}&{\! \bar M_{\nu }\simeq \left(\begin{array}{ccc}

 e& ~~c  & ~~d
\\
 c &~~ b^2 & ~~ab
 \\
d& ~~ ab  &~~ a^2
\end{array}\right)}~.
\end{array}  \!\!  ~~~
\label{nu-matrix}
\eeq
Note that  $2-3$ block's  determinant of the matrix (\ref{nu-matrix}) is also zero:
$\bar M_{\nu }^{(2,2)}\bar M_{\nu }^{(3,3)}-(\bar M_{\nu }^{(2,3)})^2= 0$.
This, using  (\ref{Ulept}), leads to the following interesting relation\footnote{See Appendix for exact expression, detailed derivation and related discussion.}
\beq
\tan^2\te_{13}\simeq \fr{m_3}{m_2}\left |s_{12}^2e^{i\rho_1}+\fr{m_2}{m_1}c_{12}^2\right | ~.
\la{nu-pred1}
\eeq
%
%
Using recent results from the neutrino experiments   \cite{nu-data, An:2012eh, Schwetz:2011zk}, we can easily verify
that the relation of Eq. (\ref{nu-pred1}) is incompatible with normal hierarchical neutrino masses.
This conclusion remains robust taking into account the effects of $1-2$ and $1-3$ rotations coming from the charged lepton sector. With these, instead of Eq. (\ref{nu-pred1}) we have the exact expression
\beq
\tan^2\te_{13}=\left |\fr{m_3}{m_2} |s_{12}^2e^{i\rho_1}+\fr{m_2}{m_1}c_{12}^2|+
\fr{K^2}{m_1m_2}e^{i\ka }\right | ~,
\la{nu-pred1-exact}
\eeq
where $K$ is real and $\ka $ is some phase. Derivation of (\ref{nu-pred1-exact}) and forms of $K, \ka $ are given in Appendix
(see Eqs. (\ref{i1})-(\ref{j})).
One can investigate for what values of $K$,
desirable values of $\te_{13}$ are obtained. In Fig. \ref{fig1}, region (i) corresponds to the values of $K$
as a function of $m_3$, which give $\te_{13}\simeq 8.9^o$. On the other hand, region (ii) shows values of $K$ obtained
within considered scenario (for $\te_{12}^e=0.016$ and $\te_{13}^e=0.0136$). We see that points of region (ii)
are well below from points of region (i).
While Fig. \ref{fig1} corresponds to the best fit values of the neutrino oscillation parameters \cite{nu-data}, the conclusion   is same by taking them within $8\si $ error bars.
This demonstrates that within considered model, the normal hierarchical neutrino mass scenario can not be realized.

\begin{figure}
\begin{center}
\leavevmode
\leavevmode
\vspace{-0.5cm}
\includegraphics{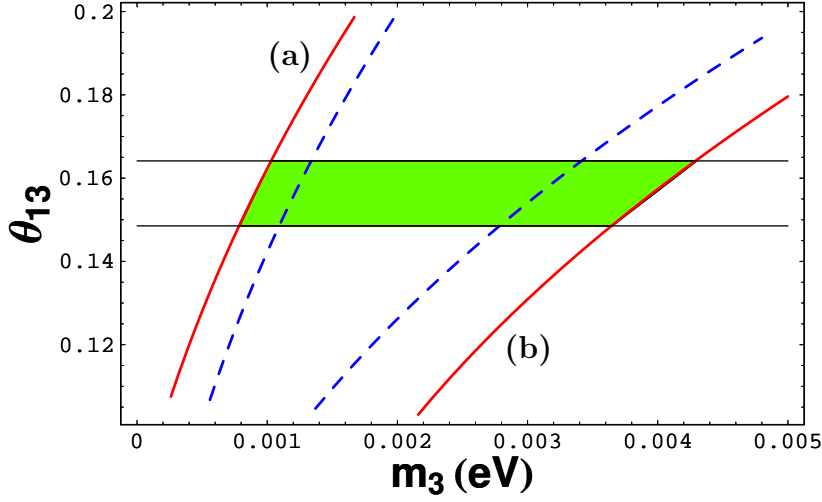}  
\end{center}
\vs{6.9cm}
\caption{Inverted hierarchical neutrino mass scenario. Green dashed region shows allowed
values of $(m_3, \te_{13})$. Red bold curves
 {\bf (a)} and  {\bf (b)} represent the dependance of $\te_{13}$ on $m_3$ for phases
 $\{\rho_1,\de+\rho_2,\om_3-\om_2,\ka \}\simeq \{0,0,0,0\}$ and $\{\pi, \pi, 0, \pi \}$ respectively.
 Curve  {\bf (a)} gives largest possible values of $\te_{13}$, while curve  {\bf (b)} -  lowest ones.
 Two horizontal lines are upper and low experimental bounds of $\te_{13}$ within the $1\si$. Dashed curves would had been
 obtained (instead of bold ones) with $K\to 0$ in Eq. (\ref{nu-pred1-exact}).}
 \rput(9.9,6){ \bf (b)}
\rput(6.8,9.9){\bf (a)}
\label{fig2}
\end{figure}

On the other hand, inverted
hierarchy in neutrino masses is possible within considered $SU(5)\tm U(1)_F$ 
model.\footnote{Worth pointing that it is usually hard to
get inverted neutrino mass scenario within GUTs \cite{Albright:2009cn}.  See however
\cite{invSO10} (within $SO(10)$) and
\cite{Stech:2008wd} (within $E_6$ GUT) with inverted hierarchical neutrino masses.}
This is demonstrated in Fig. \ref{fig2}. Green dashed region includes points captured by two border bold curves
(obtained via Eq. (\ref{nu-pred1-exact}) within our model) and two horizontal lines (corresponding to the experimental
values of $\te_{13}$ within $1\si $). This figure corresponds to the best fit values of $\te_{12}, \te_{23}, \De m_{\rm sol}^2$
and $\De m_{\rm atm}^2$, while free phases (see Eqs. (\ref{*}), (\ref{j})) are varied within full ranges.
Dashed lines correspond to the case with $K\to 0$. Thus, inclusion of the charged lepton sector somewhat extents the
allowed region. All this demonstrates
that inverted hierarchical scenario is easily realized. Fig. \ref{fig2} shows that, the allowed region for $m_3$ is fixed as:
\beq
0.0008~{\rm eV}\stackrel{<}{_\sim }m_3\stackrel{<}{_\sim }0.0044~{\rm eV}~,
\la{predictions}
\eeq
and using (\ref{inv-masses}) we get:
$$
m_1\simeq 0.04852~{\rm eV}\tm \l1+\l\fr{m_3}{0.04852~{\rm eV}}\r^2 \r^{1/2}~,
$$
\beq
m_2\simeq 0.0493~{\rm eV}\tm \l1+\l\fr{m_3}{0.0493~{\rm eV}}\r^2 \r^{1/2}~.
\la{nu-m12}
\eeq
These imply $\sum m_i\approx 0.1$~eV, satisfying the current bound \cite{Zhao:2012xw} obtained from
cosmology. Moreover, for neutrino less double $\bt $-decay
parameter $m_{\bt \bt}=|\sum U_{ei}^2m_i|$ we obtain:
\beq
m_{\bt \bt}\simeq \left |c_{12}^2m_1+s_{12}^2m_2e^{-i\rho_1} \right |~,
\la{2bt-decay}
\eeq
leading to:
\beq
0.011~{\rm eV}\stackrel{<}{_\sim }m_{\bt \bt}\stackrel{<}{_\sim }0.05~{\rm eV}~.
\la{predictions1}
\eeq
Future experiments will be able to test viability of this scenario \cite{Rodejohann:2012xd}.

\vs{0.2cm}
In summary, we have presented supersymmetric $SU(5)$ GUT supplemented with non-anomalous $U(1)_F$ flavor symmetry.
Anomaly cancellation condition restricted the field content (dictated three families of $10+\bar 5$ matter), as well as
$U(1)_F$ charge assignment. Texture zero Yukawa matrices gave natural understanding of hierarchies between charged fermion
mass and mixings. Model automatically involves $SU(5)$ singlet states utilized as right handed neutrinos. Inverted
hierarchical neutrino mass scenario is predicted within considered model. Other phenomenological issues, such as doublet-triplet
splitting, proton decay etc., left beyond the scope of this paper,  will be addressed elsewhere within more general class of models \cite{inprep} supplemented by anomaly free $U(1)_F$ symmetry \cite{Tavartkiladze:2011ex}.

\subsubsection*{Acknowledgments}

I am grateful to C.H. Albright and K.S. Babu for useful discussions and comments.
 I thank the Center for Theoretical Underground Physics and Related Areas (CETUP* 2012) for its hospitality
  and for partial support in Lead, South Dakota, where part of this work was done.
The partial support from Shota Rustaveli National Science Foundation (contract \#03/79) is kindly acknowledged.

\section*{Appendix: Effects of $\te_{ij}^e$ on the Neutrino Sector}

In this appendix we work out the details of contributions from the charged lepton sector to the
neutrino sector. In particular, as was pointed out in Sect. \ref{nu-sect}, we study impact of $\te_{12}^e$ and $\te_{13}^e$
mixing angles. With this study, we prove that within considered
SUSY $SU(5)\tm U(1)_F$ scenario, only inverted hierarchical neutrino mass scenario is realized.

Charged lepton mass terms $e^TM_Ee^c$ get diagonalized by transformations $e=L_e^*e^{\hs{0.3mm}_{'}},~e^c=R_e{e^c}'$, where $L_e, R_e$ are unitary matrices such that
\beq
L_e^\dag M_ER_e=M_E^{\rm Diag}~.
\la{ME-diag}
\eeq
Let us rotate the neutrino states $\nu$ by the same unitary transformation as $e$-states: $\nu =L_e^*\nu^{\hs{0.3mm}_{'}}$.
With this, the weak current remains diagonal: $\bar e\ga_{\mu}\nu=\bar e^{\hs{0.3mm}_{'}}\ga_{\mu}\nu^{\hs{0.3mm}_{'}}$.
On the other hand, the neutrino mass couplings $\fr{1}{2}\nu^TM_{\nu}\nu$ become
$\fr{1}{2}\nu^{\hs{0.3mm}_{'}T}\bar M_{\nu}\nu^{\hs{0.3mm}_{'}}$ with
\beq
\bar M_{\nu}=L_e^\dag M_{\nu}L_e^*~.
\la{barMnu}
\eeq
Upon transformation $\nu^{\hs{0.3mm}_{'}}=U\nu^{\hs{0.3mm}_{''}}$, the neutrino couplings can be diagonalized, i.e.
\beq
U^T\bar M_{\nu}U=M_{\nu}^{\rm Diag}~,
\la{Mnu-diag}
\eeq
and finally the weak current will be $\bar e^{\hs{0.3mm}_{'}}\ga_{\mu}U\nu^{\hs{0.3mm}_{''}}$.
Thus, the matrix $U$ in (\ref{Mnu-diag}) coincides with the lepton mixing matrix.

From Eqs. (\ref{barMnu}) and (\ref{Mnu-diag}) we obtain
\beq
L_e^\dag M_{\nu}L_e^*=U^*M_{\nu}^{\rm Diag}U^\dag ~.
\la{LMnu-UMnuDiag}
\eeq
Unitary matrix $L_e$ can be written as
\beq
L_e=P_1^lL_{23}L_{13}L_{12}P_2^l~,
\la{Le}
\eeq
where $P_{1,2}^l$ are some diagonal phase matrices, $L_{12}$, $L_{23}$ and $L_{13}$ correspond to the
rotation angles $\te_{12}^e$, $\te_{23}^e$ and $\te_{13}^e$ respectively. Without loss of generality, the
unitary matrices  $L_{12}$ and $L_{23}$ can be taken to be real orthogonal matrices. Since within our scenario
$\te_{12}^e$ and $\te_{13}^e$ are small, we can write
\beq
L_{13}L_{12}\simeq P_l'(1+\Ga )P_l''~,
\la{L13L13-Ga}
\eeq
 where $\Ga $ is real:
\beq
 \Ga= \left(\begin{array}{ccc}

 0& ~~s_{12}^{\hs{0.2mm}e}  & ~~s_{13}^{\hs{0.2mm}e}
\\
 -s_{12}^{\hs{0.2mm}e}  &~~ 0 & ~~0
 \\
-s_{13}^{\hs{0.2mm}e} & ~~ 0  &~~ 0
\end{array}\right)
~,~~~{\rm with}~~~s_{12}^{\hs{0.2mm}e}\equiv \sin \te_{12}^e~,~~~~s_{13}^{\hs{0.2mm}e}\equiv \sin \te_{13}^e~.
\label{Ga}
\eeq
Without restricting any generality, we can take $P_2^l=P_l^{''*}$ and using (\ref{Le}), (\ref{L13L13-Ga}) in
(\ref{LMnu-UMnuDiag}), we obtain
\beq
(1+\Ga^T)\hat M_{\nu}^*(1+\Ga)=UM_{\nu}^{\rm Diag}U^T~,
\la{f}
\eeq
with
\beq
\hat M_{\nu}=P_l^{'*}L_{23}^\dag P_1^{l*}M_{\nu}P_1^*L_{23}^*P_L^{'*}~.
\la{f1}
\eeq
Because of smallness of $s_{12}^{\hs{0.2mm}e}$ and  $s_{13}^{\hs{0.2mm}e}$, further we use approximation and keep first powers
of these angles (and thus first powers of the matrix $\Ga$). With this, from (\ref{f}) we get
\beq
\hat M_{\nu}^*+\Ga^T\hat M_{\nu}^*+\hat M_{\nu}^*\Ga=T~,~~~~~{\rm with}~~~~~T\equiv UM_{\nu}^{\rm Diag}U^T~.
\la{g}
\eeq
Note, that since $2-3$ block's determinant of $M_{\nu}$ is zero, similar applies to the $2-3$ block of the matrix $\hat M_{\nu}$
(see Eq. (\ref{f1})). Thus, $\hat M_{\nu}^*$ can be parameterized as
\beq
\hat M_{\nu}^* = \left(\begin{array}{ccc}

 \hat e& ~~\hat c & ~~\hat d
\\
\hat c  &~~ \hat b^2 & ~~\hat a\hat b
 \\
\hat d & ~~ \hat a\hat b  &~~\hat a^2
\end{array}\right)~.
\label{hatM-param}
\eeq
With this, using matrix relation in Eq. (\ref{g}), we derive
$$
\hat{e}-2(\hat cs_{12}^{\hs{0.2mm}e}+\hat ds_{13}^{\hs{0.2mm}e})=T_{11}~,~~~
\hat c+\hat es_{12}^{\hs{0.2mm}e}-\hat b(\hat bs_{12}^{\hs{0.2mm}e}+\hat as_{13}^{\hs{0.2mm}e})=T_{12}=T_{21}~,
$$
$$
\hat{b}^2+2\hat cs_{12}^{\hs{0.2mm}e}=T_{22}~,~~~~~~
\hat d+\hat es_{12}^{\hs{0.2mm}e}-\hat a(\hat bs_{12}^{\hs{0.2mm}e}+\hat as_{13}^{\hs{0.2mm}e})=T_{13}=T_{31}~,
$$
\beq
\hat{a}^2+2\hat ds_{13}^{\hs{0.2mm}e}=T_{33}~,~~~~~~~~~
\hat a\hat b+\hat cs_{13}^{\hs{0.2mm}e}+\hat ds_{12}^{\hs{0.2mm}e}=T_{23}=T_{32}~.
\la{h}
\eeq
By iteration (keeping ${\cal O}(s_{12}^{\hs{0.2mm}e})$ and ${\cal O}(s_{13}^{\hs{0.2mm}e})$) we obtain from (\ref{h}):
$$
\hat a^2=T_{33}-2s_{13}^{\hs{0.2mm}e}T_{13}~,~~~~~\hat b^2=T_{22}-2s_{12}^{\hs{0.2mm}e}T_{12}~,
$$
\beq
\hat a\hat b=T_{23}-s_{12}^{\hs{0.2mm}e}T_{13}-s_{13}^{\hs{0.2mm}e}T_{12}~.
\la{h1}
\eeq
These three expressions, by eliminating $\hat a$ and $\hat b$, give the following relation:
\beq
T_{23}^2-T_{22}T_{33}=2s_{12}^{\hs{0.2mm}e}(T_{23}T_{13}-T_{12}T_{33})+2s_{13}^{\hs{0.2mm}e}(T_{23}T_{12}-T_{22}T_{13})~.
\la{h11}
\eeq
Substituting $T_{ij}$ elements (see Eq. (\ref{g})) in (\ref{h11}) we obtain
\beq
m_1m_2(U_{21}U_{32}\hs{-1mm}-\hs{-1mm}U_{22}U_{31})^2\hs{-1.6mm}+
m_1m_3(U_{21}U_{33}\hs{-1mm}-\hs{-1mm}U_{23}U_{31})^2\hs{-1.6mm}+m_2m_3(U_{22}U_{33}\hs{-1mm}-\hs{-1mm}U_{23}U_{32})^2
\hs{-1.6mm}=\hs{-1.2mm}
-2s_{12}^{\hs{0.2mm}e}K_1\hs{-1.3mm}-\hs{-1.3mm}2s_{13}^{\hs{0.2mm}e}K_2
\la{i}
\eeq
with
$$
K_1\hs{-0.8mm}=\hs{-0.5mm}\fr{1}{2}\l m_1m_2\sin \hs{-0.8mm}2\te_{13}s_{23}e^{i(\de+\rho_2)}\hs{-1.4mm}+\hs{-0.5mm}
(m_1e^{i\rho_1}\hs{-1.3mm}-\hs{-1mm}m_2)m_3\sin \hs{-0.8mm}2\te_{12}c_{23}
\r \hs{-0.5mm}e^{-i(\rho_1+\rho_2+\om_1+\om_2+2\om_3)} ,
$$
\beq
K_2\hs{-0.8mm}=\hs{-0.5mm}\fr{1}{2}\l m_1m_2\sin \hs{-0.8mm}2\te_{13}c_{23}e^{i(\de+\rho_2)}\hs{-1.4mm}-\hs{-0.5mm}
(m_1e^{i\rho_1}\hs{-1.3mm}-\hs{-1mm}m_2)m_3\sin \hs{-0.8mm}2\te_{12}s_{23}
\r \hs{-0.5mm}e^{-i(\rho_1+\rho_2+\om_1+2\om_2+\om_3)}~.
\la{i1}
\eeq
Using the form of $U$ of Eq. (\ref{Ulept}) in left hand side of (\ref{i}), after some simplifications we obtain
\beq
-\tan^2\te_{13}e^{i(2\de+\rho_2)}=\fr{m_3}{m_2}\l s_{12}^2e^{i\rho_1}\hs{-1.3mm}+\hs{-0.5mm}\fr{m_2}{m_1}c_{12}^2\r \hs{-1.3mm}+\hs{-0.5mm}
\fr{2}{m_1m_2c_{13}^2}\l s_{12}^{\hs{0.2mm}e}K_1\hs{-1.3mm}+\hs{-0.8mm}s_{13}^{\hs{0.2mm}e}K_2\r
e^{i(\rho_1+\rho_2+2\om_2+2\om_3)}~.
\la{*}
\eeq
Introducing notations
$$
K^2=\fr{2}{c_{13}^2}\left | s_{12}^{\hs{0.2mm}e}K_1\hs{-1.3mm}+\hs{-0.8mm}s_{13}^{\hs{0.2mm}e}K_2\right |,~~
$$
\beq
\ka =\rho_1+\rho_2+2\om_2+2\om_2+{\rm Arg}\l s_{12}^{\hs{0.2mm}e}K_1\hs{-1.3mm}+\hs{-0.8mm}s_{13}^{\hs{0.2mm}e}K_2\r
-{\rm Arg}\l m_1s_{12}^2e^{i\rho_1}+m_2c_{12}^2\r ,
\la{j}
\eeq
from  (\ref{*}) we get Eq. (\ref{nu-pred1-exact}) - the expression for $\tan^2\te_{13}$.

Having  (\ref{*}), we can now examine possibilities of realizing normal and inverted hierarchical neutrino mass
scenarios within our model.

\subsection*{Excluding normal hierarchical neutrino mass scenario}

Let us first see if normal hierarchical neutrino masses are possible. In this case, $m_3>m_2>m_1$ and observed
mass squire differences are $\De m_{\rm sol}^2=m_2^2-m_1^2$ and $\De m_{\rm atm}^2=m_3^2-m_2^2$. Thus, two masses, say
$m_1$ and $m_2$, can be expressed as
\beq
m_1=\sq{m_3^2-\De m_{\rm sol}^2-\De m_{\rm atm}^2}~,~~~~~~~m_2=\sq{m_3^2-\De m_{\rm atm}^2}~.
\la{norm-masses}
\eeq
While  $\De m_{\rm sol}^2=m_2^2-m_1^2$, $\De m_{\rm atm}^2=m_3^2-m_2^2$ are both measured, the $m_3$ is unknown yet and we will treat it as a free parameter. Taking into account (\ref{i}), we can easily verify that $m_1=0$ is excluded. Thus,
$m_3>\sq{\De m_{\rm sol}^2+\De m_{\rm atm}^2}\simeq 0.05$~eV. On the other hand, we can also have an upper bound for $m_3$,
set from the cosmological bound on a sum of three neutrino masses
$\sum m_i \stackrel{<}{_\sim }1$~eV \cite{Zhao:2012xw, Rodejohann:2012xd}.
This, taking into account (\ref{norm-masses}), gives $m_3\stackrel{<}{_\sim }0.34$~eV.
Therefore, we will vary $m_3$ in a range
\beq
0.05~{\rm eV}\stackrel{<}{_\sim }m_3\stackrel{<}{_\sim }0.34~{\rm eV}~.
\la{norm=m3-range}
\eeq
With help of (\ref{nu-pred1-exact}) and using the best fit values of quantities $\De m_{\rm sol}^2$, $\De m_{\rm atm}^2$
and neutrino mixing angles \cite{nu-data}, we can see what values of $K$ are needed.
In Fig. \ref{fig1}, dashed region (i) represents such values of $K$ (versus $m_3$). For fixed value of
$m_3$, the multiple values of $K$ are obtained because of free phases appearing in  (\ref{nu-pred1-exact}).
On the other hand, with (\ref{j}) and (\ref{i1}) within our model with $s_{12}^{\hs{0.2mm}e}=0.016$,
$s_{13}^{\hs{0.2mm}e}=0.0136$ we can calculate $K$ for different
$m_3$ and remaining phases. Region (ii) of Fig. \ref{fig1} corresponds to this. We see that regions (i) and (ii)
do not overlap and therefore conclude that normal hierarchical neutrino mass scenario is not realized within considered model.
 This conclusion remains robust even varying
the values of $\De m_{\rm sol}^2$, $\De m_{\rm atm}^2$, $\te_{12}, \te_{23}, \te_{13}$ within
$8\si $ error bars.

\subsection*{Compatibility with inverted hierarchical neutrino masses}

As turns out, the inverted hierarchical neutrino masses blend well with relation (\ref{nu-pred1-exact}). In this case,
$\De m_{\rm sol}^2=m_2^2-m_1^2$ and $\De m_{\rm atm}^2=m_2^2-m_3^2$ and thus:
\beq
m_1=\sq{m_3^2+\De m_{\rm atm}^2-\De m_{\rm sol}^2}~,~~~~~~~m_2=\sq{m_3^2+\De m_{\rm atm}^2}~.
\la{inv-masses}
\eeq
If we set $K\to 0$ in (\ref{nu-pred1-exact}), we can easily see that for certain values of $m_3$ and $\rho_1$ all observable
can be obtained within experimentally preferred ranges. Inclusion of $K$ do not change this positive result, but
just offers slightly different choices of $m_3$ and various phases. For illustration see Fig. \ref{fig2}, with
corresponding discussion starting in a paragraph right before Eq. (\ref{predictions}).

\bibliographystyle{unsrt}

\end{document}